\documentclass[a4paper,12pt,final]{article}

\usepackage{graphicx} 
\usepackage{lineno}

\RequirePackage{xspace}
\RequirePackage{relsize}

\def\VFD{\ensuremath{V_{\text{FD}}}\xspace}

\newcommand{\tev}{\ensuremath{\mathrm{\,Te\kern -0.1em V}}\xspace}
\newcommand{\gev}{\ensuremath{\mathrm{\,Ge\kern -0.1em V}}\xspace}
\newcommand{\mev}{\ensuremath{\mathrm{\,Me\kern -0.1em V}}\xspace}
\newcommand{\kev}{\ensuremath{\mathrm{\,ke\kern -0.1em V}}\xspace}
\newcommand{\gevc}{\ensuremath{{\mathrm{\,Ge\kern -0.1em V\!/}c}}\xspace}
\newcommand{\mevc}{\ensuremath{{\mathrm{\,Me\kern -0.1em V\!/}c}}\xspace}
\newcommand{\gevcc}{\ensuremath{{\mathrm{\,Ge\kern -0.1em V\!/}c^2}}\xspace}
\newcommand{\gevgevcccc}{\ensuremath{{\mathrm{\,Ge\kern -0.1em V^2\!/}c^4}}\xspace}
\newcommand{\mevcc}{\ensuremath{{\mathrm{\,Me\kern -0.1em V\!/}c^2}}\xspace}

\def\cm   {\ensuremath{\rm \,cm}\xspace}

\def\mm   {\ensuremath{\rm \,mm}\xspace}
\def\um   {\ensuremath{\rm \,\mu m}\xspace}

\usepackage{amsmath,amssymb} 
\usepackage[per-mode=symbol]{siunitx}
\usepackage{booktabs}
\usepackage{threeparttable}
\usepackage{physics}
\usepackage{here}
\usepackage{url} 

\usepackage[normalem]{ulem}
\usepackage{xcolor}
\definecolor{mydarkgreen}{rgb}{0,0.5,0}
\definecolor{mydarkyellow}{rgb}{0.7,0.7,0}

\newcommand{\degC}{\degreeCelsius}
\newcommand{\SrSource}{${}^{90}\mathrm{Sr}\ $}

\begin{document}

\begin{center}    
{\Huge \par Radiation damage study of Belle~II silicon\\\vspace*{2mm} strip sensors with 90~MeV electron irradiation}
\\\vspace*{1.0cm}
K.~Adamczyk$^{q}$, H.~Aihara$^{o}$, K.~Amos$^{l}$, S.~Bacher$^{q}$, S.~Bahinipati$^{d}$, J.~Baudot$^{c}$, P.~K.~Behera$^{e}$, S.~Bettarini$^{i,j}$, L.~Bosisio$^{l}$, A.~Bozek$^{q}$, F.~Buchsteiner$^{a}$, G.~Casarosa$^{i,j}$, C.~Cheshta$^{t}$, L.~Corona$^{j}$, S.~B.~Das$^{f}$, G.~Dujany$^{c}$, C.~Finck$^{c}$, F.~Forti$^{i,j}$, M.~Friedl$^{a}$, A.~Gabrielli$^{j}$, V.~Gautam$^{d}$, B.~Gobbo$^{l}$, K.~Hara$^{p,m}$, T.~Higuchi$^{n}$, C.~Irmler$^{a}$, A.~Ishikawa$^{p,m}$, M.~Kaleta$^{q}$, A.~B.~Kaliyar$^{a}$, K.~H.~Kang$^{s}$, R.~Kumar$^{g}$, K.~Lalwani$^{f}$, J.~Libby$^{e}$, V.~Lisovskyi$^{b}$, L.~Massaccesi$^{i,j}$, G.~B.~Mohanty$^{h}$, S.~Mondal${\rm^{A,}}$$^{i,j}$, K.~R.~Nakamura$^{p,m}$, Z.~Natkaniec$^{q}$, Y.~Onuki$^{o}$, F.~Otani$^{n}$, A.~Paladino${\rm^{B,}}$$^{i,j}$, V.~Raj$^{e}$, K.~Ravindran$^{h}$, J.~U.~Rehman$^{q}$, I.~Ripp-Baudot$^{c}$, G.~Rizzo$^{i,j}$, Y.~Sato$^{p}$, C.~Schwanda$^{a}$, J.~Serrano$^{b}$, T.~Shimasaki$^{n}$, S.~Tanaka$^{p,m}$, R.~Thalmeier$^{a}$, T.~Tsuboyama$^{p}$, Y.~Uematsu${\rm^{C,}}$$^{o}$, L.~Vitale$^{k,l}$, S.~J.~Wang$^{o}$, Z.~Wang$^{o}$, J.~Wiechczynski$^{q}$, H.~Yin$^{a}$, L.~Zani$^{r}$, and F.~Zeng$^{n}$\hspace*{5mm}(Belle~II SVD Collaboration)
\\\vspace*{0.5cm}
{\small
$^a${\it Institute of High Energy Physics, Austrian Academy of Sciences, 1050 Vienna, Austria}\\
$^b${\it Aix Marseille Univ, CNRS/IN2P3, CPPM, Marseille, France}\\
$^c${\it IPHC, UMR 7178, Universit$\acute{e}$ de Strasbourg, CNRS, 67037 Strasbourg, France}\\
$^d${\it Indian Institute of Technology Bhubaneswar, Bhubaneswar 752050, India}\\
$^e${\it Indian Institute of Technology Madras, Chennai 600036, India}\\
$^f${\it Malaviya National Institute of Technology Jaipur, Jaipur 302017, India}\\
$^g${\it Punjab Agricultural University, Ludhiana 141004, India}\\
$^h${\it Tata Institute of Fundamental Research, Mumbai 400005, India}\\
$^i${\it Dipartimento di Fisica, Universit\`{a} di Pisa, I-56127 Pisa, Italy}\\
$^j${\it INFN Sezione di Pisa, I-56127 Pisa, Italy\\ ${^{A}}$Presently at INFN Sezione di Torino, I-10125 Torino, Italy\\ $^{B}$Presently at INFN Sezione di Bologna, I-40127 Bologna, Italy}\\
$^k${\it Dipartimento di Fisica, Universit\`{a} di Trieste, I-34127 Trieste, Italy}\\
$^l${\it INFN Sezione di Trieste, I-34127 Trieste, Italy}\\
$^m${\it The Graduate University for Advanced Studies (SOKENDAI), Hayama 240-0193, Japan}\\
$^n${\it Kavli Institute for the Physics and Mathematics of the Universe, University of Tokyo, Kashiwa 277-8583, Japan}\\
$^o${\it Department of Physics, University of Tokyo, Tokyo 113-0033, Japan\\ $^{C}$Presently at High Energy Accelerator Research Organization (KEK), Tsukuba 305-0801, Japan}\\
$^p${\it High Energy Accelerator Research Organization (KEK), Tsukuba 305-0801, Japan}\\
$^q${\it H. Niewodniczanski Institute of Nuclear Physics, Krakow 31-342, Poland}\\
$^r${\it INFN Sezione di Roma Tre, I-00185 Roma, Italy}\\
$^s${\it Department of Physics, Kyungpook National University, Daegu 41566, Korea}\\
$^t${\it Dr. B.R. Ambedkar National Institute of Technology, Jalandhar, Punjab 144008, India}}\\
\vspace*{0.8cm}
{August 27, 2025}
\ \vspace*{0.8cm}
\end{center}

\newpage

\begin{abstract}

The silicon strip sensors of the Belle~II silicon vertex detector were irradiated with \SI{90}{MeV} electron beams up to
a 1~MeV neutron equivalent fluence of $3.1\times 10^{13}~\si{cm^{-2}}$, which corresponds to approximately ten years of the full-luminosity Belle~II operation, including a safety factor of greater than three in the background estimation.  Electrons with comparable energies are one of the major sources of beam-induced background in the silicon strip sensors during Belle~II operation.
We measure radiation-induced changes in sensor properties, including the full depletion voltage, sensor leakage current, noise, and charge collection.

For the sensors developed for the Belle~II silicon vertex detector, the sensor bulk
effective doping changes polarity at
a 1~MeV neutron equivalent fluence of $(6\mbox{-}7)\times 10^{12}$~\si{cm^{-2}}.
We also determine a damage constant of $3.7\times 10^{-17}$ A/\cm at \SI{17}{\degC} immediately after irradiation.
Due to annealing, the damage constant decreases to about 40\% of the initial value at a temperature of approximately 17°C in the first 200 hours. The temperature is then set to 20°C between 200 and 400 hours, and to 25°C after 400 hours; as a result, the decrease stabilizes at around 30\% of the initial value after 1000 hours.
We measure sensor noise and signal charge for a sensor irradiated with
{the 1~\mev neutron-equivalent fluence} of ${3.1}\times 10^{13}$~\si{cm^{-2}}.  Noise increases by approximately 44\% after irradiation,
while {charge collection} does not change significantly
when a sufficiently high bias voltage is applied.

Even after radiation damage corresponding to a fluence of \SI{3.1e13}{\centi\meter^{-2}}, these measurements confirm that the full depletion voltage and leakage current will remain within the acceptable range of the power supply system, and that a sufficient signal-to-noise ratio can still be maintained despite the increase in noise.

\end{abstract}


\section{Introduction}
The Belle~II experiment~\cite{Belle-II:2010dht} searches for new particle phenomena beyond the Standard Model of particle physics in collisions of \SI{7}{\giga\electronvolt} electrons and \SI{4}{\giga\electronvolt} positrons, provided by the SuperKEKB {collider}~\cite{Ohnishi:2013fma} with a target integrated luminosity of \SI{50}{\atto\barn^{-1}}.
{The Belle~II detector must operate in a high-background environment induced by the high-intensity beams.}

The silicon vertex detector (SVD)~\cite{Belle-IISVD:2022upf}, together with the pixel detector, composes the inner tracking system of the Belle~II detector.
Due to beam-induced background particles, we expect a challenging radiation environment for the SVD.
The estimated total ionizing radiation dose and {Non-Ionizing Energy Loss (NIEL)} in
{1~MeV neutron equivalent fluence}
on the innermost SVD sensors {for ten years of the full-luminosity Belle~II operation are about $\SI{35}{kGy}$ and ${8\times 10^{12}}$~\si{cm^{-2}}}, respectively~{\cite{SVD:AnnualDose, Belle-IIBeamBG:2023upf}}. These estimates come from a beam-background simulation that assumes that SuperKEKB operates with its target luminosity of $6 \times 10^{35} \cm^{-2}\text{s}^{-1}$.
{One of the} major sources of radiation damage to SVD sensors is electrons in the energy range of 1 to \SI{100}{\mega\electronvolt} based on the beam background studies~\cite{Belle-IIBeamBG:2023upf}.
{The radiation damage during Belle~II SVD operation causes an increase in leakage current, a change in the full depletion voltage, an increase in noise, and a possible change in the charge-collection efficiency.}
Therefore, the long-term operation of Belle~II requires an understanding of the radiation tolerance and behavior of the irradiated sensors.

We performed an irradiation test of the SVD sensors using \SI{90}{\mega\electronvolt} electron beams produced at the Research Center for Electron Photon Science (ELPH), Tohoku University, Sendai, Japan~{\cite{RARiS}}.
{The sensors were irradiated up to
{a 1~MeV neutron equivalent fluence}
of \SI{3.1e13}{\centi\meter^{-2}}, and the radiation-damage effects in their characteristics, including the leakage current, full depletion voltage, noise, and charge collection, were measured in order to investigate whether the effects of radiation damage on the sensor and detector performance remain acceptable under the Belle~II SVD operating conditions.}
{Throughout the experiment, the leakage currents were measured in a dark environment.}

The remainder of the paper is structured as follows. 
Section \ref{sec:setup} describes the experimental setup, including irradiated samples, an overview of the irradiation test, and measurement methods.
Section \ref{sec:results_and_discussions} discusses the results of the measurements. Finally, Section \ref{sec:conclusion} presents the conclusions.

\section{Setup}
\label{sec:setup}
\subsection{Irradiated samples and sensor assemblies}
\begin{figure}[htpb]
 \centering
   \includegraphics[width=\linewidth]{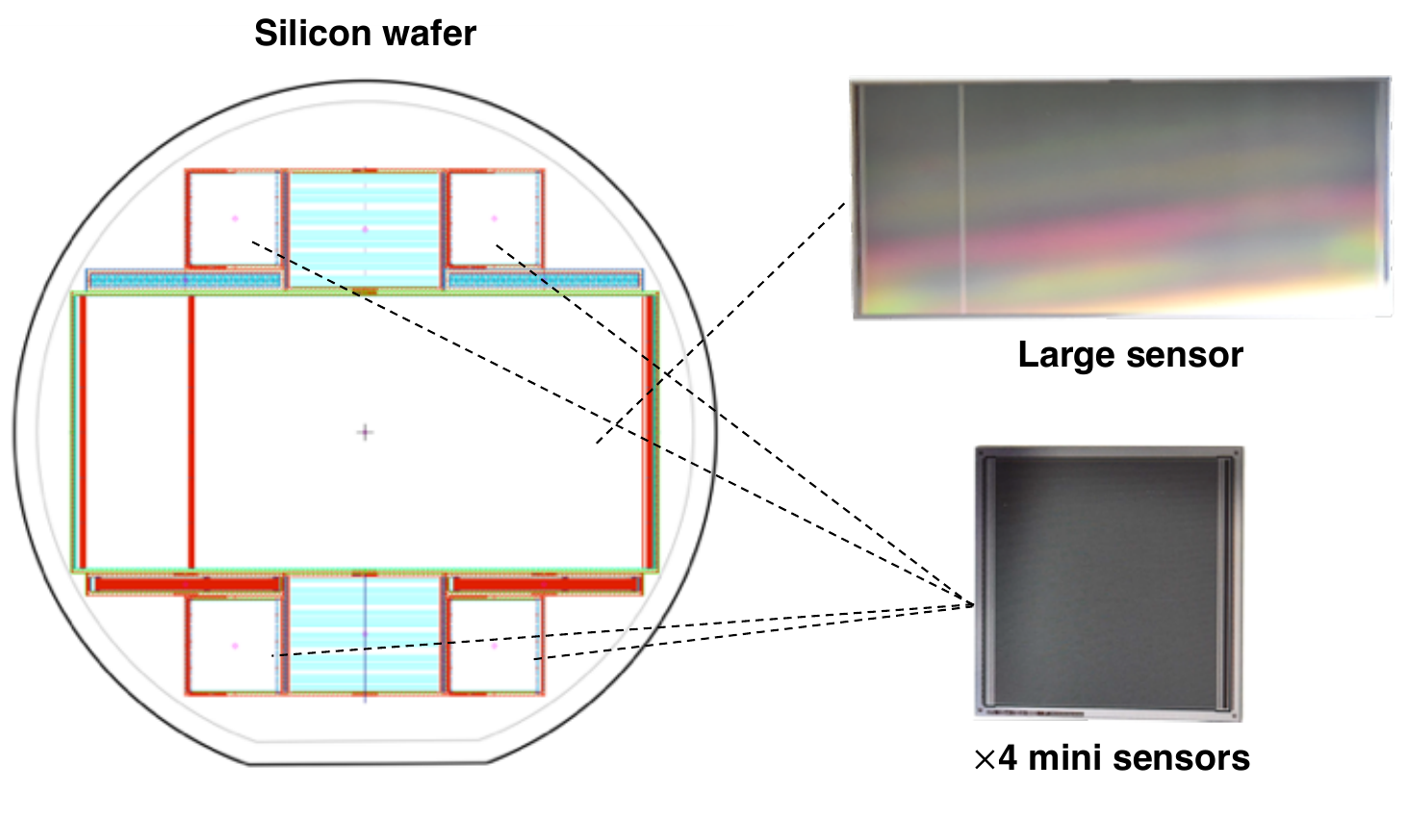}
    \caption{
    {Layout of one large and four mini sensors on a {150 mm} diameter wafer, as viewed with the P-side facing up.}
    {The P-strips on both sensors are aligned parallel to the longitudinal axis of the large sensor. The N-strips are perpendicular to this axis on the opposite side of the wafer.}
    }
\label{fig:wafer}
\end{figure}

We irradiated two different types of double-sided silicon-strip sensors, here called large sensors and mini sensors. These sensors were produced by Hamamatsu Photonics K.K., {Hamamatsu, Japan}. 
{The large sensor is one of the large rectangular sensors fabricated for the Belle~II SVD~\cite{Belle-IISVD:2022upf}, and its quality is the same as that of the installed sensors.}
{The mini sensors were fabricated together with the large sensors as the test elements located in extra space on {150 mm diameter} silicon wafers as shown in Figure~\ref{fig:wafer}.}
Therefore, {these sensors share the same bulk properties}.
{The large and the mini sensors are fabricated on \SI{320}{\micro\meter} thick N-type wafers with a resistivity of \SI{8}{\kilo\ohm\centi\meter}.}
{The full depletion voltage is less than \SI{120}{\volt}. The breakdown voltage is well above the full depletion voltage, by more than \SI{50}{\volt}.}

The opposite sensor faces have implanted strips heavily doped with either acceptors or donors.
The side with acceptors is called the P-side, whereas the side with donors is called the N-side.
The strips on the P- and N-sides are referred to as P- and N-strips, respectively.
{To improve the position resolution, the large sensor has one intermediate floating strip, not connected to a readout channel, positioned between each pair of adjacent readout strips on both P and N-sides.}
The mini sensors do not have floating strips.
The readout strips are AC-coupled to aluminum strips, from which they are separated by a silicon-oxide dielectric layer.
Table~\ref{tab:sensors} lists the geometrical parameters of the large and mini sensors. 
The parameters of the large sensor are identical to those used in the Belle~II SVD, whereas those of the small sensors are different except for the sensor thickness.

\begin{table}[htbp]
\caption{Geometrical parameters of irradiated sensor samples.}\label{tab:sensors}
\centering
\begin{threeparttable}
\begin{tabular}{lrrr}
    \toprule
    & \multicolumn{1}{r}{Large sensor} & \multicolumn{1}{r}{Mini sensor} \\
    \midrule
    Thickness [\si{\um}] & 320 & 320 \\ 
    Sensor size [$\si{\cm\times\cm}$]& $12.492\times5.964$ & $2.12\times2.12$ \\
    Sensitive area [$\si{\cm\times\cm}$]& $12.29\times5.772$ & $1.92\times1.92$\\
    Number of strips P-side/N-side & 768/512\tnote{*} & 192/192\\
    Strip pitch P-side/N-side [\si{\um}] & 75/240\tnote{*} & 100/100 \\ 
    Strip length P-side/N-side [\si{\cm}] & 12.29/5.772 & 1.92/1.92 \\ 
    \bottomrule
\end{tabular}
\centering
\begin{tablenotes}\footnotesize
     \item{*} The large sensor has readout strips and
     floating strips.  The quoted values of number and pitch are for readout strips only.
\end{tablenotes}
\end{threeparttable}
\end{table}

\begin{figure}[htpb]
     \centering
     \includegraphics[width=0.65\textwidth]{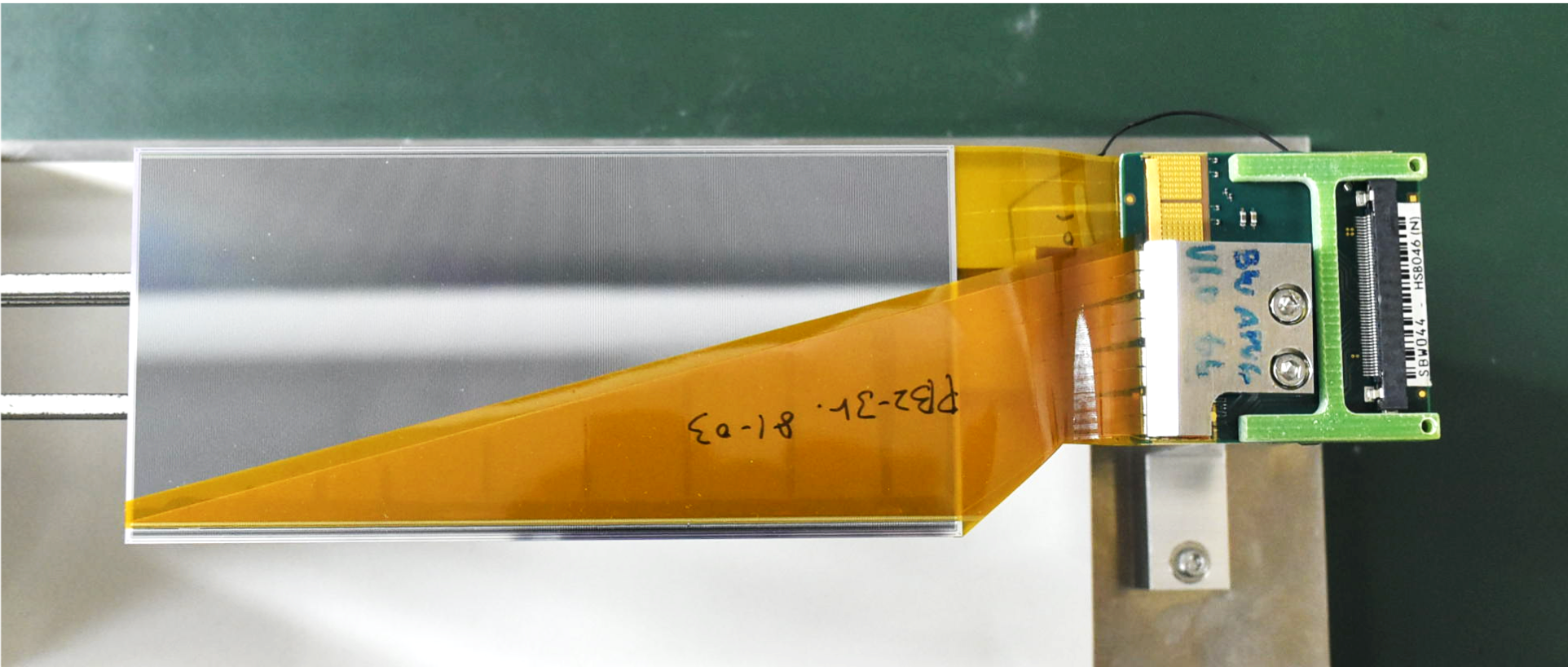}
     \caption{{The large sensor integrated with the {front-end readout} components, including a flexible circuit and APV25 readout chips.}}
     \label{fig:SBW}
\end{figure}

We irradiated two large sensors and ten mini sensors.
One of the two large sensors, labeled ``large-1'', and the ten mini sensors, labeled ``mini-$n$'' with $n$ ranging from 1 to 10, were
individually mounted on a printed circuit board (PCB) that
allowed application of a bias voltage via wire-bonded connections to the sensor bias pads {for all strips to bias commonly through poly resistors}. 

The other large sensor was connected via a flexible circuit~\cite{Irmler:2016qyg} to a readout circuit board equipped with APV25 readout chips~\cite{French:2001xb}, which enabled us to measure noise and collected charge.
{The module is shown in Figure~\ref{fig:SBW} and} is referred to as the test-module in the paper. 
The APV25 chip samples the detector signals with a \SI{32}{MHz} sampling clock.
{The APV25 chip is confirmed to have sufficient radiation tolerance, and its performance is not affected at the radiation levels anticipated for the Belle~II SVD.}
The flash ADC readout system digitized the APV25 output for signal reconstruction; the same system is used for the SVD readout~\cite{Belle-IISVD:2022upf}.

\subsection{Beam facility and irradiation setup}\label{sec:beamfa}

The injector LINAC of the booster synchrotron ring at ELPH provided \SI{90}{\mega\electronvolt} electron beams.
The maximum beam current {in the experiment} was \SI{140}{\nA}, and the beam-spot shape was a circle with a diameter of approximately \SI{1}{\cm}.
{The beam current was reduced to \SI{20}{\nA} for the lower irradiation samples by reducing the pulse repetition frequency from \SI{7}{\Hz} to \SI{1}{\Hz}}.

\begin{figure}[htpb]
    \centering
    \includegraphics[width=\linewidth]{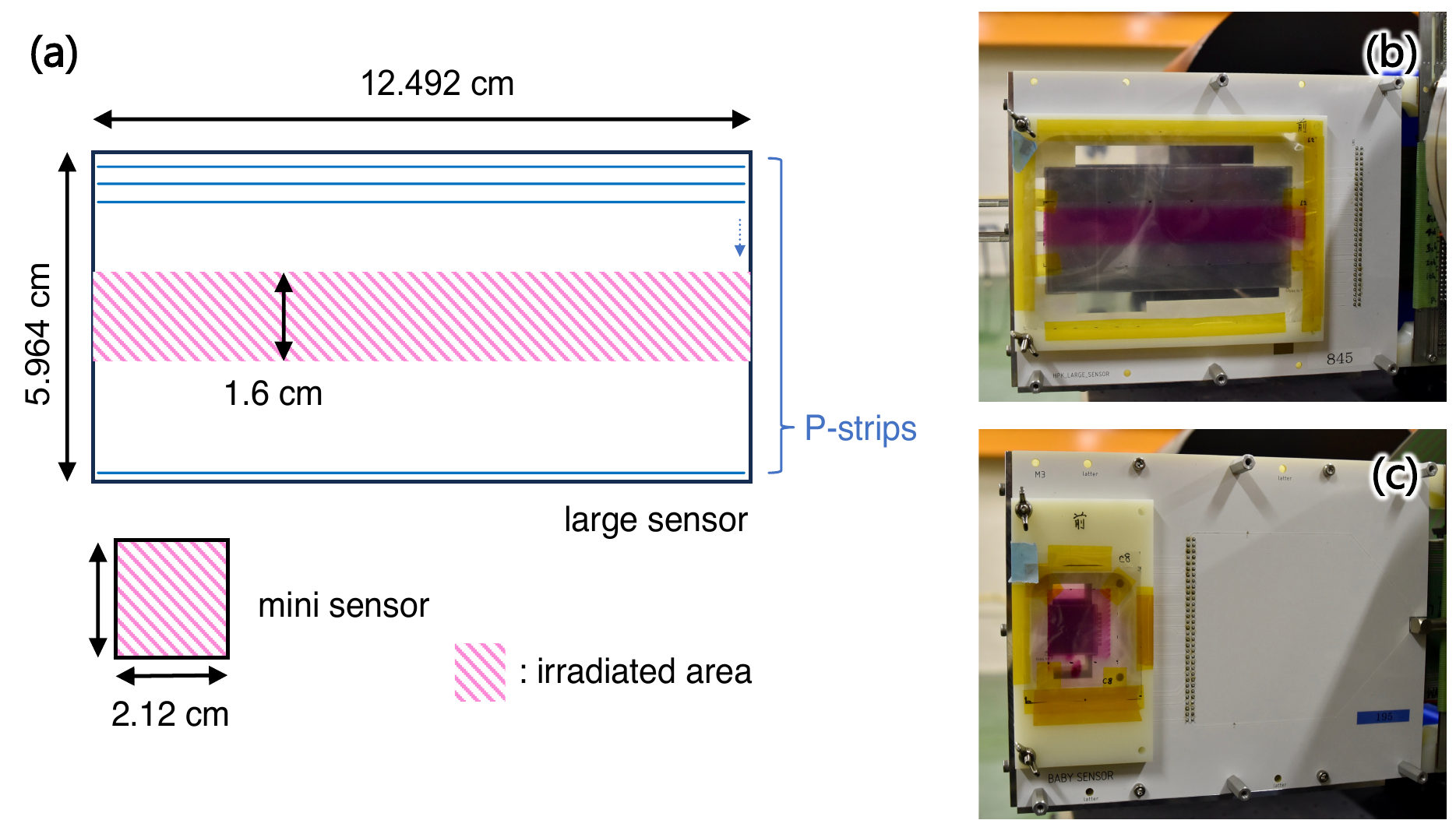}
    \caption{
    {(a):} Schematic layout of the irradiated area {for the large sensor and the mini sensor. (b) and (c):} Irradiated large and mini sensor, respectively, each mounted on a PCB. The irradiated regions are identified by the areas where the attached radiochromic film on the sensor turned magenta. 
    In this study, we did not use the films for the dose estimation.}
    \label{fig:irradiated_area}
\end{figure}

To correctly evaluate the radiation-damage effect on strip noise and charge collection, the entire sensitive area under the strips must be irradiated uniformly.
We aligned the two mini sensors with the beamline and uniformly irradiated them by moving them in the plane perpendicular to the beamline using two motorized linear stages.  
The movement speed was set to {\SI{1.0}{\cm\per\s}}.
Figure~\ref{fig:irradiated_area} illustrates the irradiated sensor areas. 
{For the mini sensors, the entire sensor area was irradiated uniformly.}
{For the large sensor, irradiation of the entire sensor area was not possible because of the limited allocated beam time. One of the main concerns after irradiation was a possible increase in P-strip noise after type inversion, caused by the larger inter-strip capacitance resulting from the change in effective silicon bulk doping from N-type to P-type. Therefore, partial irradiation was performed on a 1.6-cm-wide region covering the full length of a subset of the P-strips, allowing the effect of irradiation on the P-strip noise to be evaluated. Since only a portion of the N-strip length was irradiated, the post-irradiation performance evaluation was carried out mainly for the P-strips.}
During the irradiation, no bias voltage was applied to {both the mini and large} sensors.

We performed the electron-beam irradiation in
multiple
steps {of increasing fluence}. 
After each step, we made an $I$-$V$ measurement with the setup described in {Section}~\ref{subsec:iv}.
{During the irradiation and the $I$-$V$ measurement, the sensor temperature was kept at approximately \SI{17}{\degC}.}
We derive the electron fluence from the electron beam current and exposure time. 
{The electron fluence for one sweep of the sensor area is $1.3 \times 10^{12}$ ($8.8\times10^{12}$)  \si{cm^{-2}} at the beam current of 20 (140) \si{\nA}, which corresponds to the radiation dose of about 0.32 (2.2) \si{\kilo\gray}.}
The {1~MeV neutron equivalent fluence}
{is obtained by multiplying} the electron fluence with the hardness factor for \SI{90}{MeV} electrons ($\sim 0.08$~\cite{HardnessFactor,RD50}).
For the mini sensor, the irradiation rate at the maximum beam current of \SI{140}{\nA} was estimated to be about \SI{29}{Gy/s} and about $9.0 \times 10^{9}$ \si{\cm^{-2}\s^{-1}} in radiation dose and {1~MeV neutron equivalent fluence}, respectively.

\begin{figure}[htpb]
    \centering
    \includegraphics[width=0.8\linewidth]{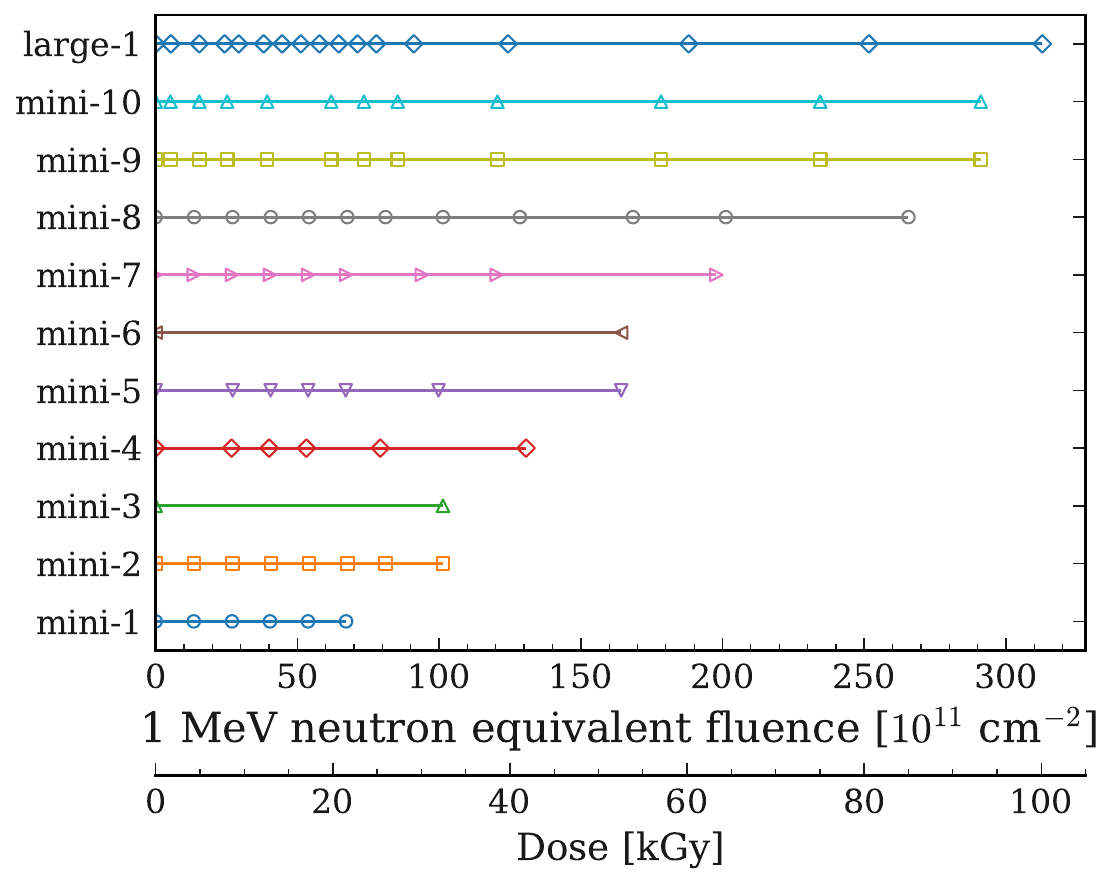}
    \caption{
    {Points in the figure represent estimated cumulative {1~MeV neutron equivalent fluence} at the timing of I-V measurement conducted.} 
    {The rightmost point represents the total radiation dose for each sensor.}
    {The uncertainty in the {1~MeV neutron equivalent fluence} values is estimated to be 6\%.}
    }
    \label{fig:dose}
\end{figure}

Figure~\ref{fig:dose} summarizes the radiation dose and
{1~MeV neutron equivalent fluence} on the sensors mounted on the PCBs.
The cumulative electron fluence varied
{from sensor to sensor} to study the effect of different levels of radiation damage.
The maximum {electron fluences (radiation doses) } on the large and mini sensors were { $3.9\times 10^{14}$ \si{\cm^{-2}} (\SI{100}{\kilo\gray}) and $3.6 \times 10^{14}$ \si{\cm^{-2}} (\SI{93}{\kilo\gray})}, which correspond to
{1~MeV neutron equivalent fluences} of
${3.1}\times 10^{13} ~ \si{cm^{-2}}$ and ${2.9}\times 10^{13}~\si{cm^{-2}}$, respectively.  {The total uncertainty in the {1~MeV neutron equivalent fluence} is estimated to be 6\%, obtained by adding in quadrature the 2\% uncertainty arising from {beam current} fluctuations and the 6\% uncertainty associated with the hardness factor.}
We irradiated the large sensor on the test-module to the same cumulative radiation dose as the large sensor mounted on the PCB.
{The radiation fluence achieved in the irradiation test is sufficient to conservatively evaluate the impact of radiation damage expected under Belle~II operating conditions; the total-fluence
value exceeds {the anticipated {1~MeV neutron equivalent fluence} of $8\times 10^{12}$~\si{\cm^{-2}}} expected for ten years of operation at full luminosity, corresponding to a safety factor of greater than three with respect to this expected value.}
{However, it should be noted that the irradiation rate in the test differs from the realistic environment, and possible rate-dependent effects are not taken into account.}

\subsection{Setup for $I$-$V$ measurements}
\label{subsec:iv}
A Keithley~2614B source meter \cite{Keithley} measured the leakage current of the sensors on the PCBs while applying a variable bias voltage. This is called an $I$-$V$ measurement.

{During the irradiation test, }{the $I$-$V$ measurements and beam irradiations were alternatively performed to track the evolution of leakage current and full depletion voltage with increasing cumulative dose on each sensor, with the exception of mini-3 and mini-6.}

To study annealing effects on the radiation damage, after irradiation 
we {stored the sensors at approximately \SI{17}{\degC} for the first 200 hours after the irradiation, \SI{20}{\degC} between 200 and 400 hours, and \SI{25}{\degC} after 400 hours, 
and} made $I$-$V$ measurements at several elapsed times since the last irradiation step. These elapsed times are approximately 30, 200, 1000, 2000, and 10000 hours. 
{For comparison, the temperature conditions of the SVD during operation and shutdown are as follows. During Belle~II operation, which typically lasts for several months each year, the SVD sensor temperatures range from approximately \SI{5}{\degC} in the outermost layer to approximately \SI{30}{\degC} in the innermost layer, depending on the sensor position, although the temperature of the innermost layer is not precisely known because the innermost-layer sensors are not equipped with temperature sensors. During the shutdown period, which occupies the remaining part of the year, the sensor temperatures remain around \SI{25}{\degC}.}

The square of the measured leakage current, $I^2$, is approximately proportional to the bias voltage $V$ {up to the full-depletion voltage $V_{\text{FD}}$, provided that the irradiation fluence remains moderate,} as  
{the leakage current is expressed as $I=q_e U S_{\text{sensor}} w$, where $w$ is the depletion depth, and the depletion depth follows $w=\sqrt{2\epsilon_{\rm{Si}}V/q_e|N_\text{eff}|}$}: 
\begin{equation}\label{eq:I2V}
    {I}^2 \simeq (q_e US_{\text{sensor}})^2\frac{2\epsilon_{\rm{Si}}V}{q_e \abs{N_\text{eff}}} \propto V \quad (V<V_{\text{FD}}),
\end{equation}
where $q_e$ is the elementary charge, $U$ is the electron-hole-pair generation rate per unit volume, $S_{\text{sensor}}$ is the sensor area, $\epsilon_{\text{Si}}$ is the silicon dielectric permittivity,  $N_\text{eff}$ is the effective carrier concentration, and $V_{\text{FD}}$ is the full depletion voltage.
{After the irradiation, the electron-hole-pair generation rate increases and the effective carrier concentration changes~\cite{Moll:1999kv}.}
Upon reaching $V\!=\!V_{\text{FD}}$, the leakage current approximately {saturates since the bulk becomes fully depleted at this voltage and the depletion volume remains almost unchanged}.
This behavior of the leakage current assumes the following: the dominant component of the leakage current is carrier generation in the depleted semiconductor bulk, and
the dopant concentration is uniform across the depth of the sensor.
Due in part to the carrier generation at the sensor surface, as well as other surface-related currents, and in part to the approximations implicit in Eq.~(\ref{eq:I2V}), 
a discrepancy from $I^2\propto V$
can be expected at high radiation dose.

By analyzing the measured $I$-$V$ curves {to find a transition voltage between the region of $I^2 \propto V$ and the saturated region}, we evaluate the full depletion voltage and the leakage current as described in {Sections}~\ref{sec:full_depletion_voltage} and  \ref{sec:LeakageCurrent}, respectively.

\subsection{Setup for noise and charge collection measurements}
\label{sec:setup_noise_cce}
\begin{figure}[htpb]
    \centering
    \includegraphics[width=0.5\linewidth]{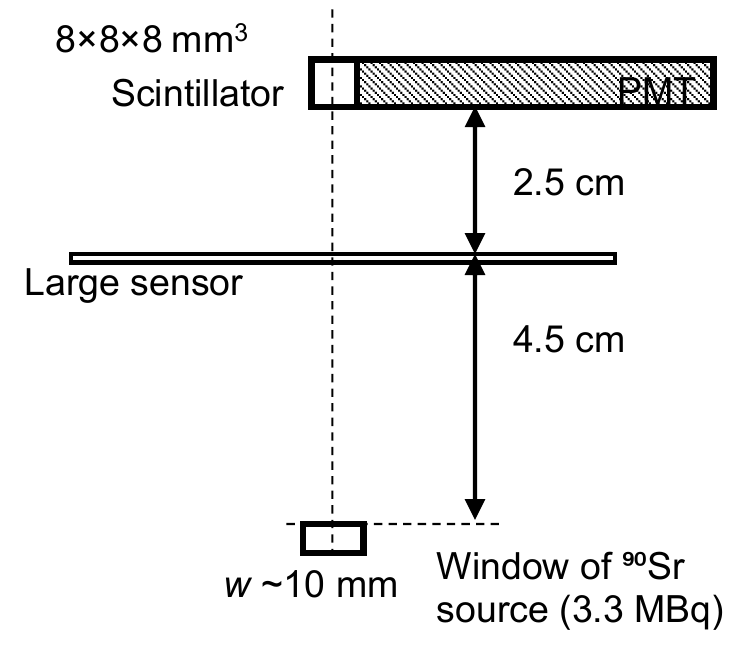}
    \caption{
        Overall configuration of the $^{90}$Sr-source test system. The test-module sensor is placed in the middle of the system. The scintillator with a dimension of $8\times 8 \times 8\,\si{\cubic\mm}$ is mounted \SI{2.5}{\cm} above the sensor, and the \SrSource source is set \SI{4.5}{\cm} below the sensor.}
    \label{fig:CCE_setup}
\end{figure}

We measured the strip noise and signal charge generated in the test-module sensor by beta-decay electrons 
from a \SI{3.3}{MBq} \SrSource source.
The maximum energy of \SrSource electrons is \SI{2.3}{\mega\electronvolt}.
The overall configuration of the $^{90}$Sr-source test system is illustrated in Figure~\ref{fig:CCE_setup}.
A scintillator, {read out with a photomultiplier tube (PMT)}, detected the electrons passing through the test-module sensor, which served as a trigger for the signal readout. We aligned the scintillator and \SrSource source with
the center of the test-module sensor.
{The sensor temperature can be considered approximately equal to room temperature \SI{24}{\degC}, and the noise was confirmed to be stable.}
{A single scintillator was used to trigger the data acquisition. In the charge-collection analysis, events were selected by requiring a cluster in the scintillator-triggered readout window, thereby rejecting accidental scintillator triggers without a corresponding sensor signal.
}

{We performed} {noise and} {charge collection measurements before irradiation and} {after the 1000-hour annealing step following irradiation, as defined in Section~\ref{subsec:iv}.}
Both measurements were performed at approximately \SI{24}{\degC}, which is close to the representative temperature {estimated for the innermost-layer SVD sensors.}
The bias voltage for the sensor increased in \SI{10}{V} steps up to \SI{100}{V} in the measurement before the irradiation and up to \SI{160}{V} after the irradiation. {The maximum applied bias voltage was increased because of the change in the full depletion voltage after the irradiation.}

In the charge-collection measurement, we use a clustering algorithm~\cite{Belle-IISVD:2022upf}, which 
groups some neighboring strips and adds charges from these strips to obtain the
{cluster charge}. 
We require the strip charge to be at least five times
the strip noise at the seed strip of a cluster and three times
the noise at adjacent strips, 
where we define a seed strip to be that with the largest signal charge.

\section{Results and Discussion}
\label{sec:results_and_discussions}
\subsection{Full depletion voltage}
\label{sec:full_depletion_voltage}

\begin{figure}
  \centering
  \begin{minipage}{0.49\textwidth}
      \centering
      \includegraphics[width=\textwidth]{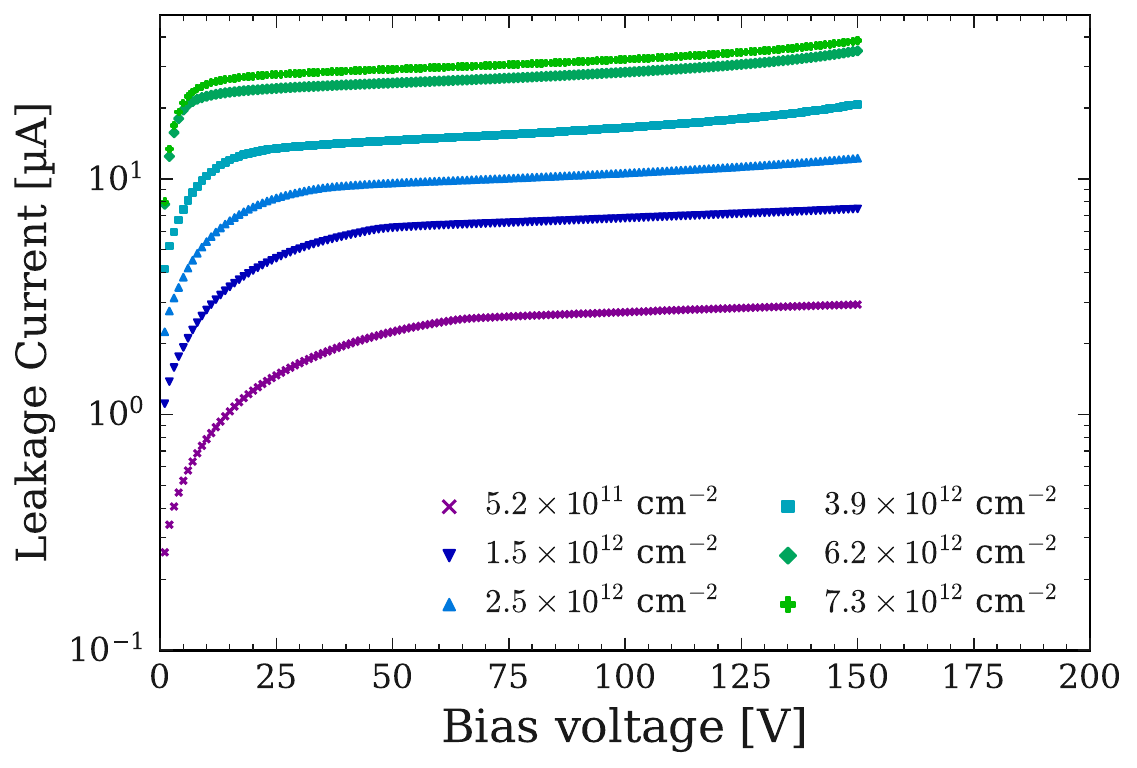}
  \end{minipage}
  \hfill
  \begin{minipage}{0.49\textwidth}
    \centering
    \includegraphics[width=\textwidth]{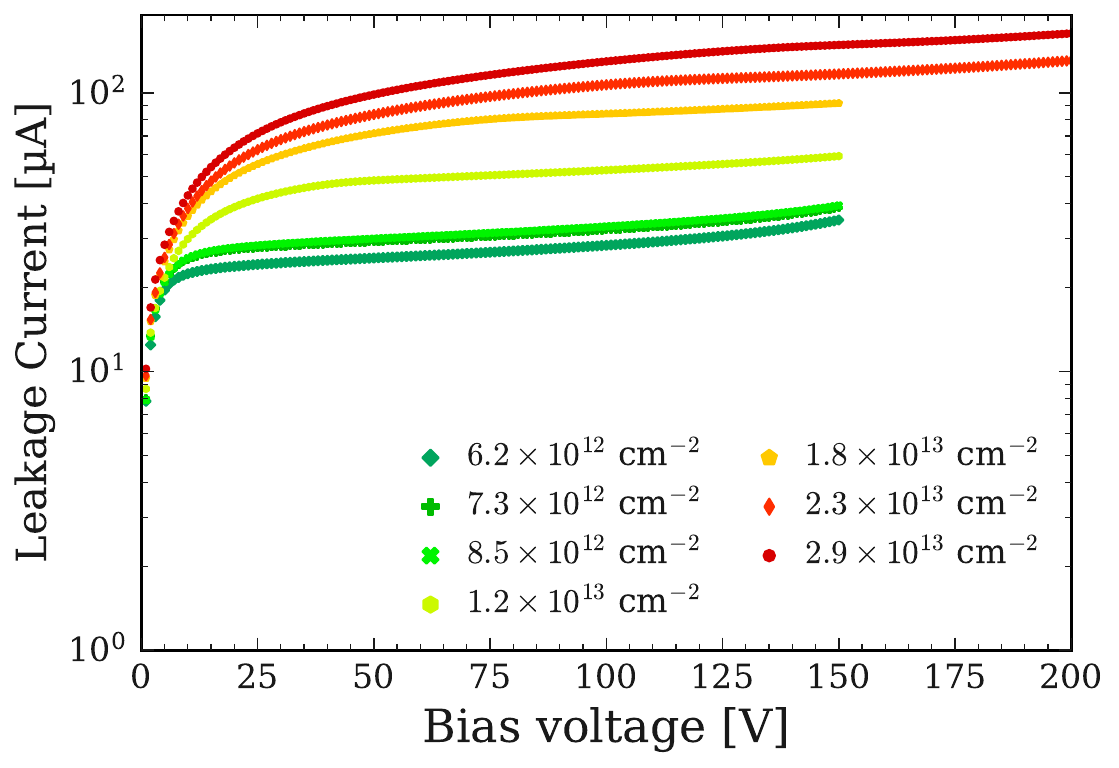}
  \end{minipage}
  \centering
  \begin{minipage}{0.49\textwidth}
    \centering
    \includegraphics[width=\textwidth]{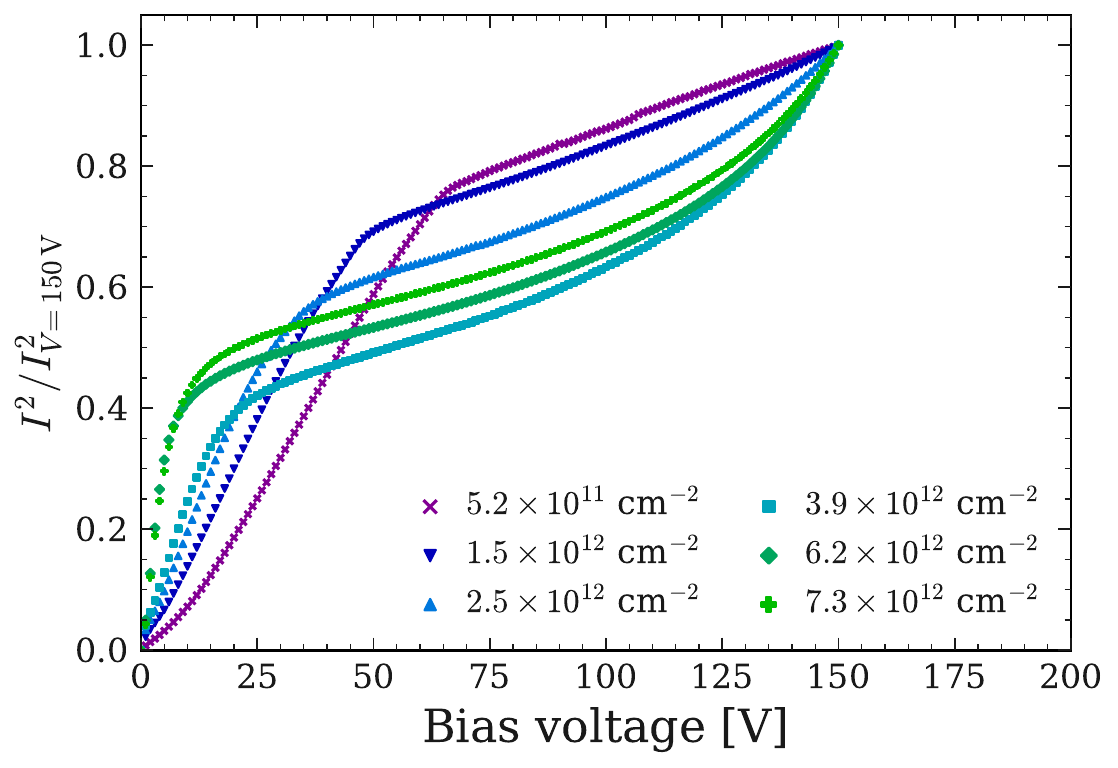}
  \end{minipage}
  \hfill
  \begin{minipage}{0.49\textwidth}
    \centering
    \includegraphics[width=\textwidth]{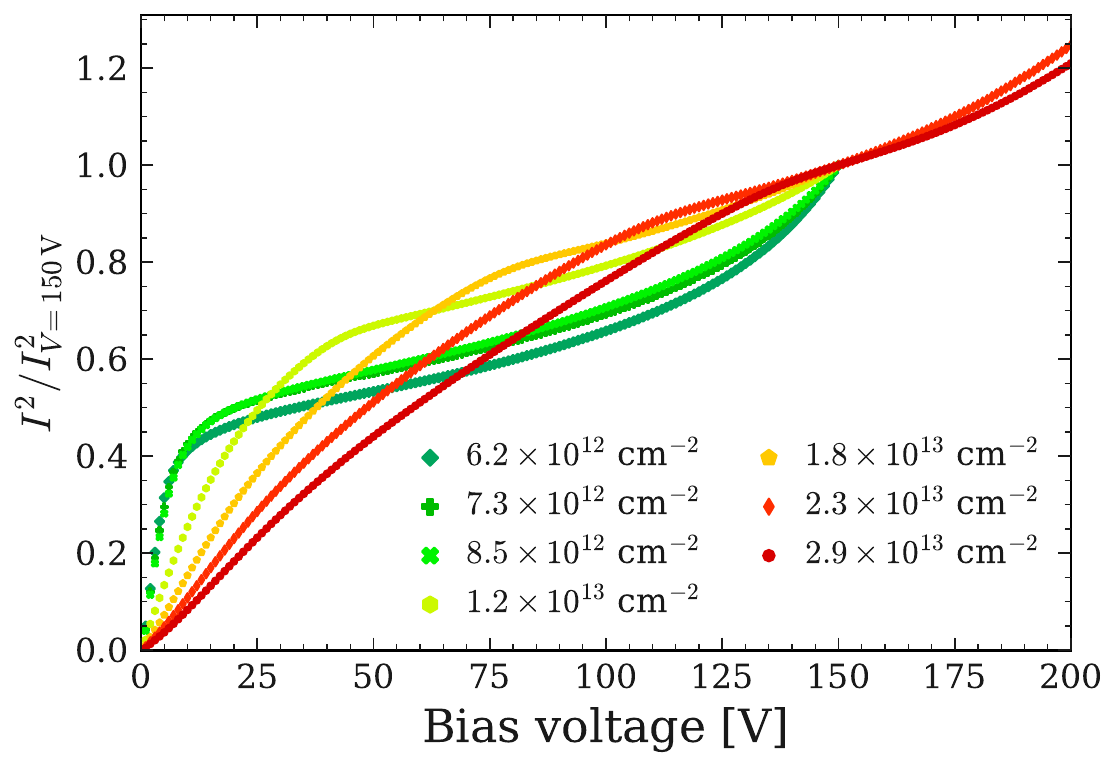}  
  \end{minipage}
  \caption{$I$-$V$ (top) and normalized $I^2$-$V$ (bottom) curves for the mini-9 sensor.
  For each {1~MeV neutron equivalent fluence}, the $I^2$-$V$ curve is normalized by the square of the leakage current at \SI{150}{\volt}. 
  The range of fluences shown is {${[5.2, 73]}\times10^{11}~\si{\cm^{-2}}$ (left) and ${[6.2, 29]\times10^{12}}~\si{\cm^{-2}}$ (right)}.
  {For ${2.3\times10^{13}}~\si{\cm^{-2}}$ and ${2.9\times10^{13}}~\si{\cm^{-2}}$, data are taken up to \SI{200}{\volt}
  because the knee of the $I^2$-$V$ curves is at a voltage higher than \SI{150}{\volt}.}
  }
  \label{fig:I2V}
\end{figure}

{As discussed in Section~\ref{subsec:iv}},
the $I^2$-$V$ curve {is expected to exhibit} a distinct knee at
$V_\text{FD}$. 
Figure~\ref{fig:I2V} shows the $I^2$-$V$ curves for the mini-9 sensor at different
{1~MeV neutron equivalent fluences},
normalized by the square of the leakage current at \SI{150}{\V}.
The location of the knee determines the full depletion voltage at the radiation dose; the definition of the location is the local minimum value of the second derivative of the $I^2$-$V$ curve. {The derivatives were calculated numerically by taking the differences between adjacent data points. A quadratic function was fitted to the region around the minimum to determine the local minimum value.}
When the full depletion voltage becomes less than \SI{20}{\volt}, the local minimum derived from the second derivative is difficult to identify due to the rapid increase in leakage current. 
In such cases, we determine the full depletion voltage from the slope values of the $I^2$-$V$ curve.
\ref{sec:logI2} provides more details about the estimation of low full depletion voltages.

\begin{figure}[htbp]
    \centering
    \includegraphics[width=\linewidth]{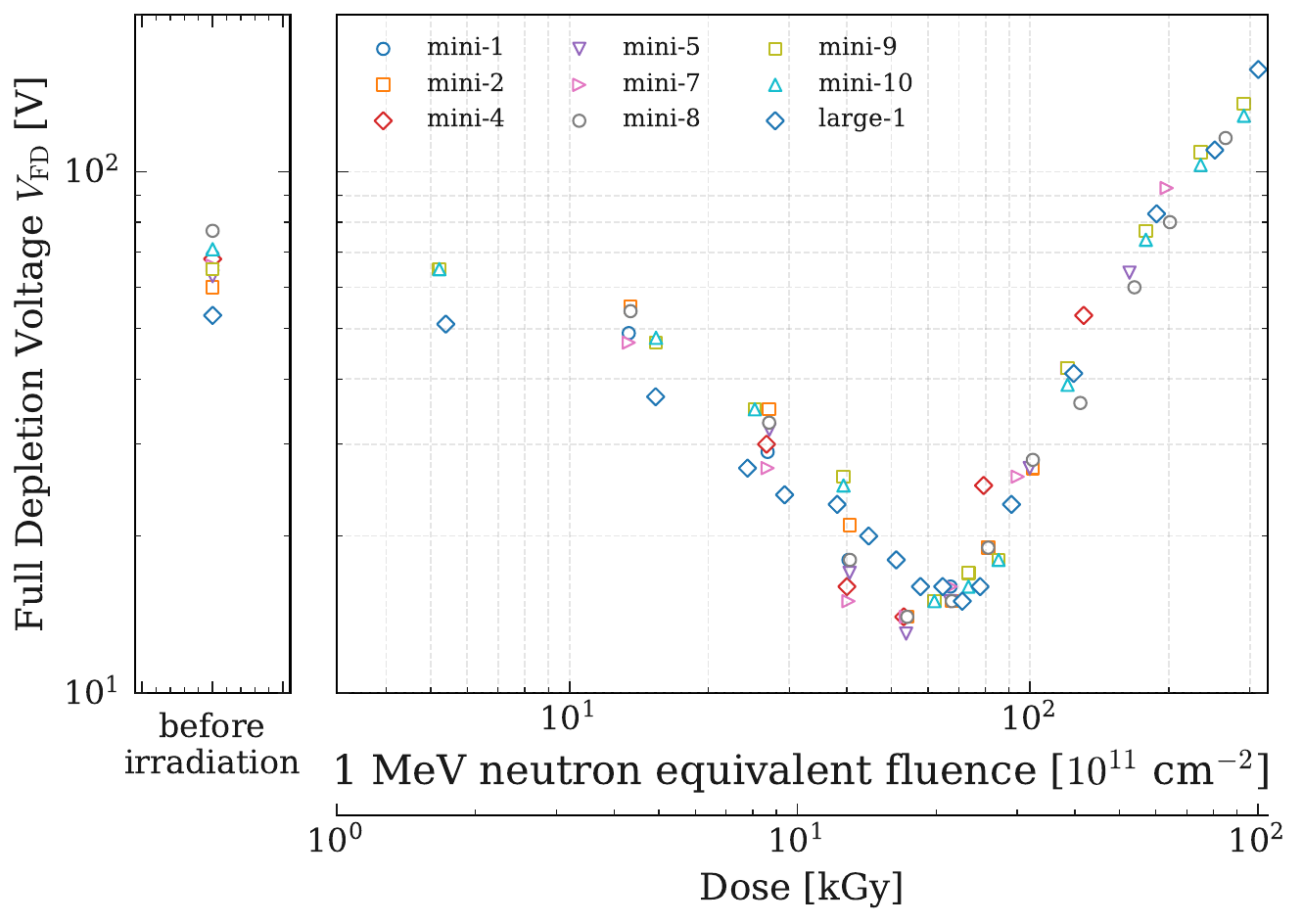}
    \caption{Variation of the full depletion voltage with {1~MeV neutron equivalent fluence} for the irradiated sensor samples.
    The points in the figure indicate the full depletion voltage measured immediately after each irradiation step.
    }
    \label{fig:TypeInversion}
\end{figure}

Figure~\ref{fig:TypeInversion} shows the evaluated full depletion voltages as a function of radiation dose.
{Type inversion is found to occur around {a 1~MeV neutron equivalent fluence} of ${(6\mbox{-}7)}\times10^{12}~\si{\cm^{-2}}$}.
{The full depletion voltage increases after the type inversion and reaches about \SI{160}{\volt} at ${3.1}\times 10^{13}$ \si{cm^{-2}}, which approximately corresponds to ten years full-luminosity operation of Belle~II with a safety factor greater than three.
This value is comfortably below the maximum applicable bias voltage (\SI{200}{\volt})
from the power supply system of the SVD~\cite{Belle-IISVD:2022upf}.
}

\begin{figure}[htbp]
    \centering
    \includegraphics[width=0.8\linewidth]{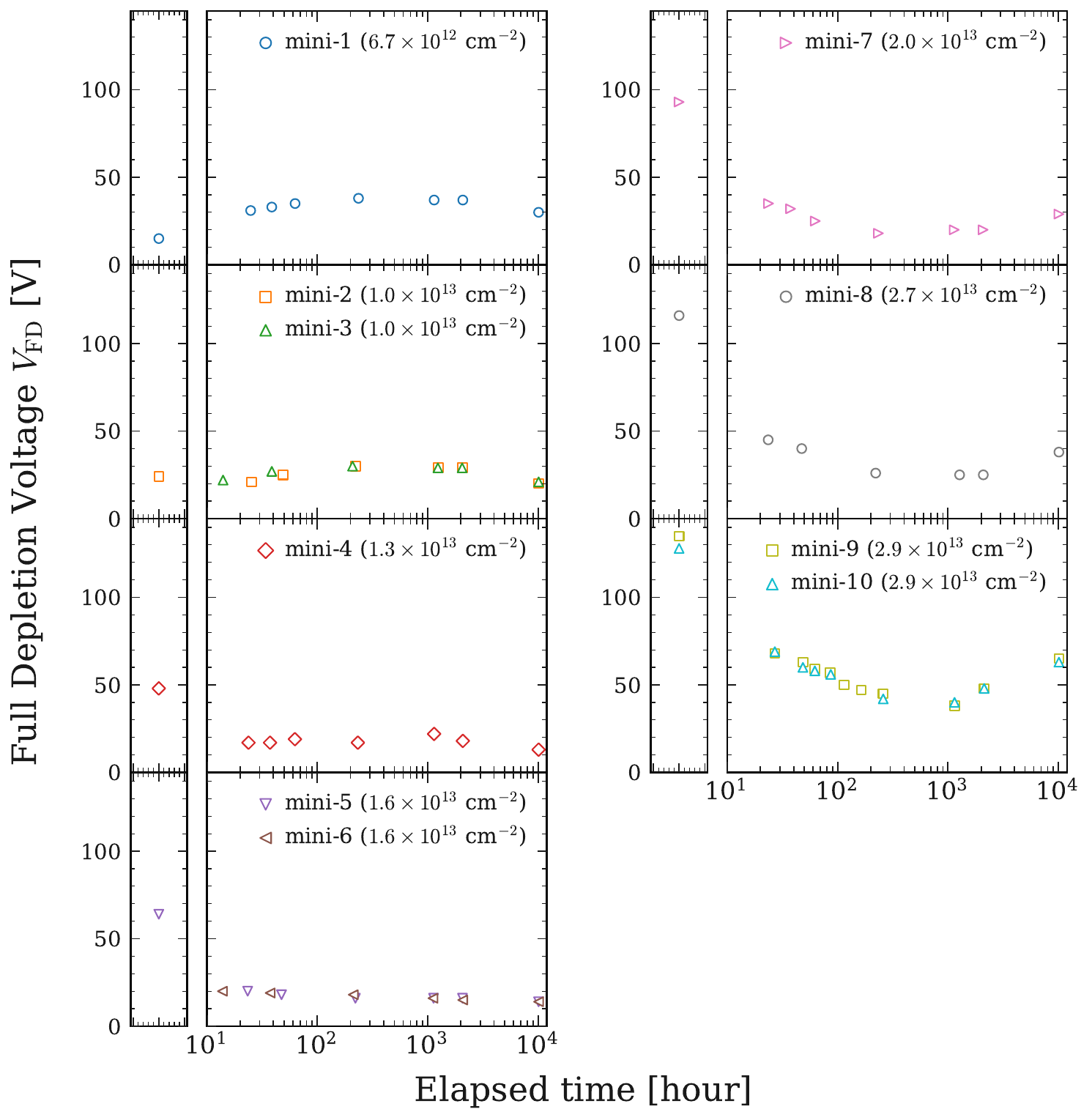}
    \caption{{The left} panels show the full depletion voltage of each irradiated sensor just after its last irradiation; the right panels show changes in the full depletion voltage of each irradiated sensor over the elapsed time after the last irradiation.}
    \label{fig:VFDvsTIME}
\end{figure}

We also measured the {annealing effect on} the full depletion voltage of mini sensors
due to the elapsed time since the end of their irradiation. 
Figure~\ref{fig:VFDvsTIME} shows the evolution of the full depletion voltage.
Sensors mini-7 to mini-10 were exposed to doses much higher than that required for type inversion. In these sensors, the observed time evolution of the total depletion voltage is consistent with the behavior expected for type-inverted sensors, according to established annealing models~\cite{Moll:1999kv,LINDSTROM2001308}.  We observed a decrease in the full depletion voltage between 200 and 1000~hours after irradiation {down to about \SI{50}{\volt}}, which is due to
beneficial annealing. This is followed by an increase after 1000 hours, consistent with the reverse-annealing effect, which is characterized by a longer time constant.  In contrast, sensors mini-1 to mini-6, which were exposed to smaller doses, showed less pronounced change in depletion voltage over time, except for the first measurement taken about 20 hours after completion of the irradiation steps.
This might be due to a reduced impact of the annealing effect, which is also expected to {be proportional to} {the 1~MeV neutron equivalent fluences}~{\cite{LINDSTROM2001308}}.

\subsection{Leakage current}
\label{sec:LeakageCurrent}
\begin{figure}[htbp]
    \centering
    \includegraphics[width=0.8\linewidth]{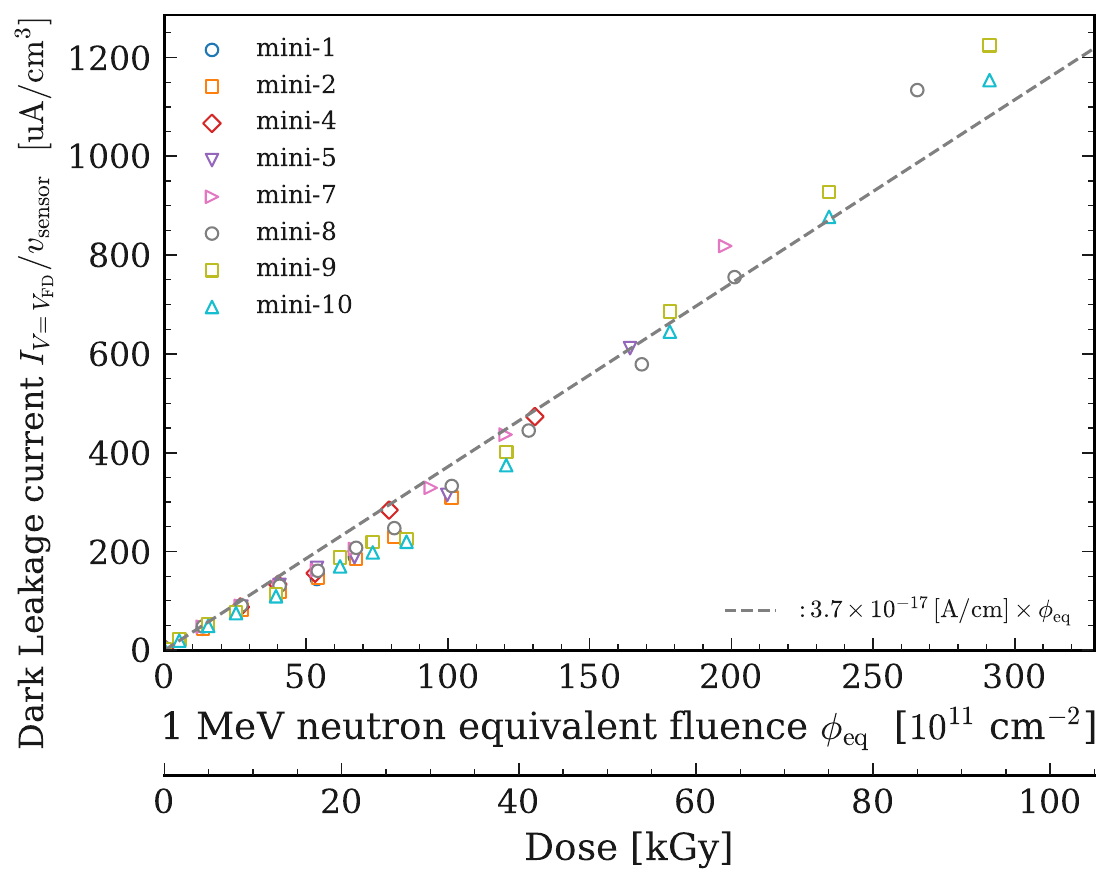}
    \caption{
    Leakage current per sensor volume ($v_\text{sensor}$) measured right after the irradiation.
    The ambient temperature at the measurement time was \SI{17}{\degC}.}
    \label{fig:DamageFactor}
\end{figure}

The bulk leakage current is known to increase proportionally with {NIEL}~\cite{Moll:1999kv,LINDSTROM2001308}.
The slope between the leakage current
{normalized to} the sensor volume {in fully depleted sensor}, $I/v_\text{sensor}$, and the neutron-equivalent fluence, $\phi_\text{eq}$, defines the damage constant.
We measured the bulk leakage current of the irradiated sensors applying the full depletion voltage, $I_{V=\VFD}$.
Figure \ref{fig:DamageFactor} shows the distribution of $I/v_\text{sensor}$, measured immediately after irradiation, as a function of $\phi_\text{eq}$.
The leakage current exhibits a linear increase with $\phi_\text{eq}$.
To determine the damage constant, we fit a straight line to the $I/v_\text{sensor}$ distribution.
The resulting damage constant at an ambient temperature of $\SI{17}{\degC}$ is 
${3.7}\times10^{-17}~{\rm A/cm}$.
Assuming that the leakage current is proportional to $T^2 \exp(-E_g/2kT)$, {where $T$ is the temperature}, $E_g$ is the silicon band gap, and $k$ is the Boltzmann constant,
we obtain for the damage constant scaled to a temperature of \SI{20}{\degC} a value of
${4.9}\times10^{-17}~{\rm A/cm}$.
{Because the annealing during the irradiation and the intervals between the irradiation steps were not sufficiently controlled and the irradiated samples were stored at three different temperatures in different time periods after irradiation, direct quantitative comparisons of the obtained damage constant with results from other studies should be interpreted with care.}

{Two independent sources are considered to be the dominant contributors to the uncertainty in the damage constant. The first arises from the uncertainty in determining the full depletion voltage, which propagates to the leakage current value evaluated at that voltage, results in an approximately 5\% relative uncertainty in the damage constant. The second originates from the choice of bias voltage used for evaluating the leakage current (e.g., using a voltage of \SI{10}{\volt} above full depletion), leading to an approximately 5\% variation in the damage constant. Assuming these two contributions are independent, the total relative uncertainty is estimated to be about 7\% by adding them in quadrature.}

\begin{figure}[htpb]
    \centering
    \includegraphics[width=0.8\linewidth]{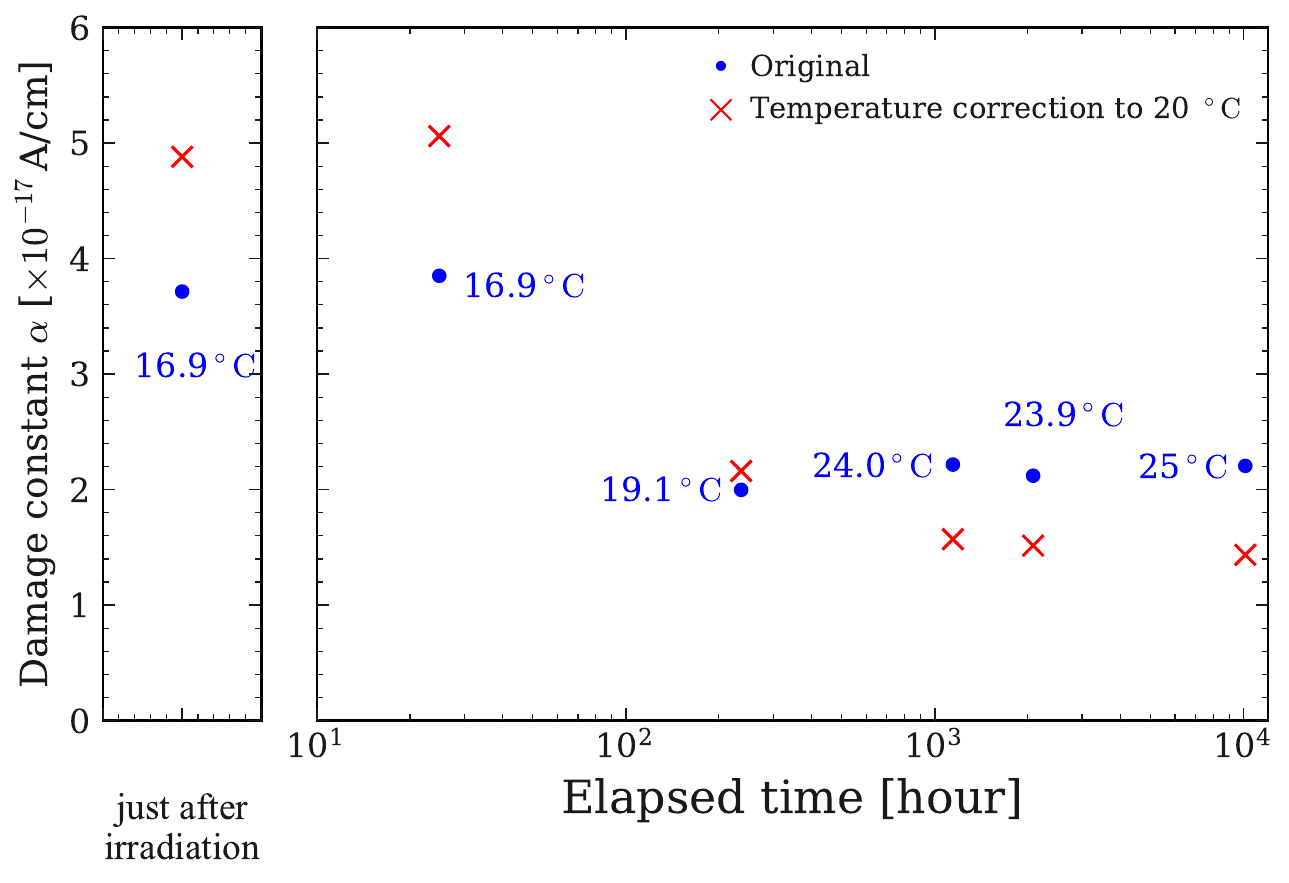} 
    \caption{{The left panel shows the damage constant just after irradiation, corresponding to Fig.~\ref{fig:DamageFactor} and is derived using leakage current measurements from all irradiation steps. The right panel shows the time evolution of the damage constant calculated from the final radiation dose and leakage current of all mini sensors obtained after the final irradiation step.} The measured results at each time point are indicated by filled dots, with the nearby numbers representing the corresponding measurement temperatures. Crosses represent temperature-scaled damage constants to \SI{20}{\degC}. }
    \label{fig:DamageFactorAnnealing}
\end{figure}

We also measured {the annealing effect on} the leakage current at various elapsed times after the irradiation finished.
Figure \ref{fig:DamageFactorAnnealing} shows the damage constant's time evolution.
The damage constant was measured at ambient temperature then scaled to \SI{20}{\degC}.
The damage constant decreases significantly to approximately $40\%$ of the initial value 200 hours after irradiation, then it decreases to about $30\%$ {of the initial value} 1000 hours after irradiation.

{The damage constant of the SVD sensors installed and operated in the Belle~II experiment was estimated from measurements of the leakage current and estimates of the radiation damage~\cite{Belle-IISVD:2022upf, YoSato:2023}. In the estimation of the radiation damage, the dose measured continuously by the surrounding diamond sensors was converted to the dose on the SVD sensors by taking into account the correlation with the SVD sensor hit occupancy. The obtained dose was then converted into the {1~MeV neutron equivalent fluence} using a dose-to-fluence conversion factor determined from simulations.}
The {damage constant} value ranges from approximately $2\times10^{-17}~{\rm A/cm}$ to $10\times10^{-17}~{\rm A/cm}$, for sensors across different layers, with the operating temperature varying from approximately \SI{5}{\degC} to approximately \SI{30}{\degC}. {Though precise comparison is difficult due to significant uncertainties from} imprecise neutron-equivalent fluence and temperature measurements {for the installed sensors, the magnitudes of these results} are consistent with the value reported here.

{The leakage current of the installed SVD sensors after radiation damage corresponding to a fluence of \SI{3.1e13}{\centi\meter^{-2}}, that is estimated from the results of this study while considering the annealing effect and possible operation temperature of the SVD sensors, remains within the range determined by the current rating of the power supply system.}

\subsection{Noise}

{As described in Section \ref{sec:setup_noise_cce}, we performed noise and charge collection measurements at the temperature of \SI{24}{\degC} after irradiation {by a 1~MeV neutron equivalent fluence of $3.1 \times 10^{13}~{\rm cm}^{-2}$} and subsequent annealing for 1000 hours, in order to compare the results with those obtained before irradiation.}
{The test module was irradiated within the area shown in Figure~\ref{fig:irradiated_area}.}
We performed the strip noise and charge collection measurements for the P-strips because the irradiation of the test-module sensor was along the P-strips.
We measure the noise value in terms of ADC counts, which is converted to the equivalent noise charge (ENC),
using the gain measured with 
the internally generated calibration pulses of the APV25 chip.
This conversion method is the same as that for the Belle~II SVD \cite{Belle-IISVD:2022upf}. 
Due to processing variations, the absolute values of the calibrated gain vary between different APV25 chips; therefore, we estimate that the accuracy of the gain measurement is about 15\%.

\begin{figure}[htpb]
    \centering
    \includegraphics[width=\linewidth]{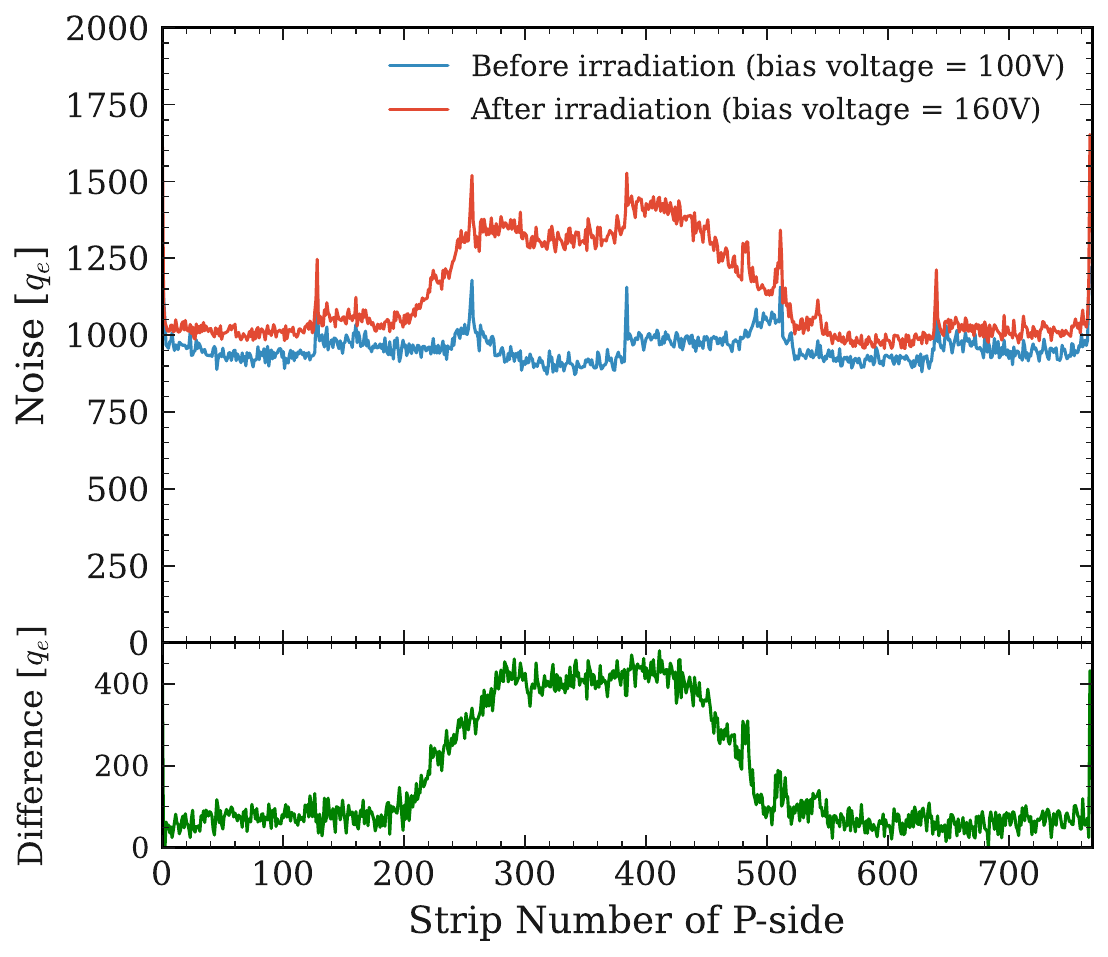}
    \caption{Noise of P-strips of the test-module sensor measured before and after the irradiation (upper panel) and their difference (lower panel).
    {The spikes observed every 128 strips (APV25 borders) are known to be caused by crosstalk originating from the power inputs of the chips.}
    }
    \label{fig:noise}
\end{figure}

Figure \ref{fig:noise} shows the P-strip noise, {expressed in units of the elementary charge $q_e$}, when applying a bias voltage of \SI{100}{\volt} before irradiation and \SI{160}{\volt} after irradiation.
We observe a significant increase in noise after irradiation in the central region of the sensor, which corresponds to the irradiated area. 
The increase is approximately 410~{$q_e$}, which corresponds to a relative increase of $44\%$.

Even in the region not irradiated by the electron beam, the noise increases by approximately 50~$q_e$. This is likely the result of a non-negligible dose of {the background radiation by particles scattered in the beam line components.}

We use the measured damage constant to estimate the contribution to ENC from the leakage current;
{Section}~\ref{sec:LeakageCurrent} describes the damage-constant measurement.
A representation of the ENC due to the leakage current is as follows \cite{Hartmann:2017gzy}:
\begin{equation}
    \mbox{ENC}/{q_e} = \frac{\mathrm{e}}{2}\sqrt{\frac{i\cdot t}{q_e}} \simeq 107 \sqrt{i/{\rm nA}\cdot t/{\rm \mu s}}
\end{equation}
where $\mathrm{e}$ is Euler's number, $i$ is the leakage current of a strip, and $t$ is the shaping time.
The damage constant, measured 1000 hours after the end of irradiation at the ambient temperature of $\SI{24}{\degC}$, is approximately
${2.2}\times10^{-17}~{\rm A/cm}$.
Considering the volume of a P-strip and the {1~MeV neutron equivalent fluence}, we find the leakage current per P-strip is \SI{2300}{\nano\ampere}.
With an APV25 shaping time of \SI{50}{\nano\second}, the resulting ENC is approximately 1100 {$q_e$}.

The total noise value in the irradiated region is 1400~{$q_e$}, as shown in Figure~\ref{fig:noise}. This value is consistent with the quadratic sum of the
noise value of 1000~{$q_e$} contributed by the inter-strip capacitance and the noise value of 1100~{$q_e$} contributed by the leakage current.
{Another effect that should be considered as a possible cause of the radiation-induced noise is surface damage. Surface damage creates fixed oxide charges, which increase the inter-strip capacitance and consequently the noise~\cite{Belle-IISVD:2022upf}. However, since no bias voltage was applied to the sensors during irradiation, the surface damage in this study possibly have been smaller than that expected under the actual SVD operating conditions, and the resulting increase in noise is therefore also expected to be smaller.}

\subsection{Signal charge collection}
For the signal charge measurement, we used P-strips in the region where the noise increased uniformly after irradiation. Figure~\ref{fig:noise} shows that these strips have IDs in the range 300 to 399.
We evaluate the difference in signal charges before and after irradiation. We fit a Landau function convoluted with a Gaussian function to the signal charge distribution; the most probable value (MPV) of the function estimates the collected charge.

\begin{figure}[htbp]
    \centering
    \includegraphics[width=\linewidth]{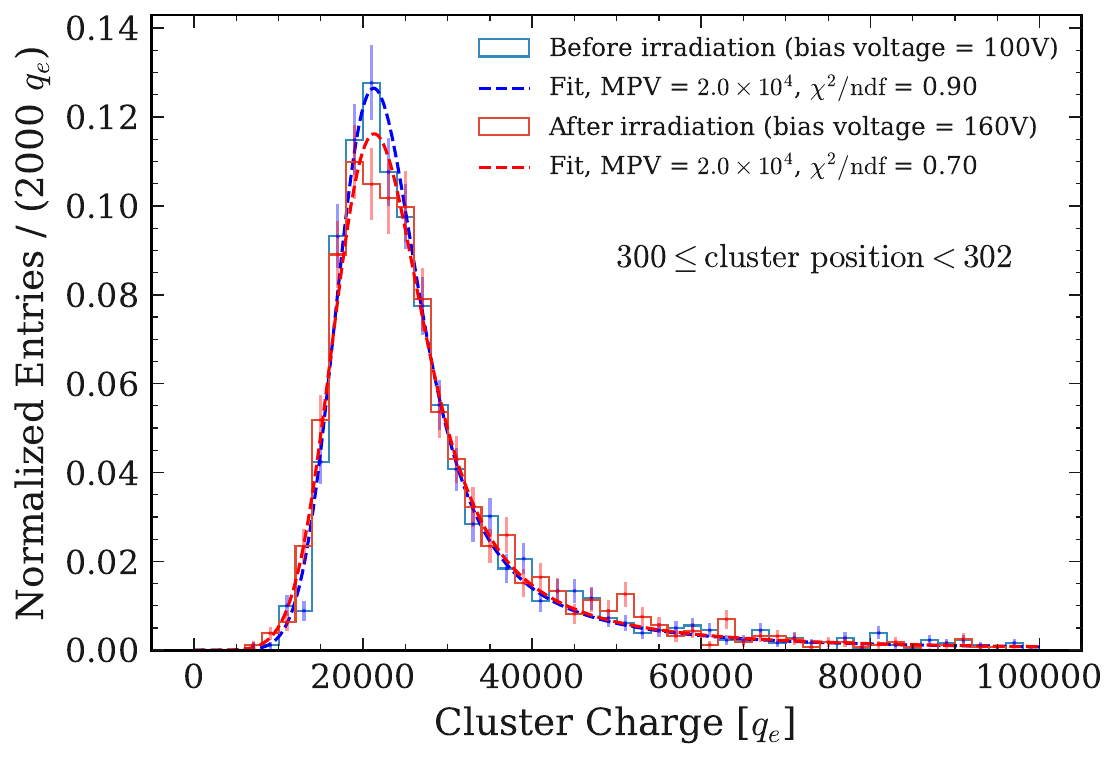}
    \caption{{The} distributions of {the signal charge}
    for P-strip clusters with positions in the range $300 \le \text{strip ID} < 302$
    before (blue) and after (red) irradiation, measured at bias voltages of $\SI{100}{\V}$ and $\SI{160}{\V}$, respectively.  Distributions are normalized by the number of {the total entries} in this region. A Landau function convoluted with a Gaussian function is fitted to the distributions.  {The reduced $\chi^2$'s of the fits are shown in the legend labels.}}
    \label{fig:CCdistribution}
\end{figure}

{To estimate the variation of the signal charge MPV, we classified the observed clusters into 50 sub-regions of the 100-strip region according to their cluster position.}
Figure~\ref{fig:CCdistribution} shows an example of a normalized distribution of the signal charge for a cluster with a position in the range $300 \le \text{strip ID} < 302$.
{Here, bias voltages of 100 V and 160 V were applied before and after irradiation, respectively, at which the MPVs had already reached saturation, as discussed below}.

\begin{figure}[htpb]
    \centering
    \includegraphics[width=\linewidth]{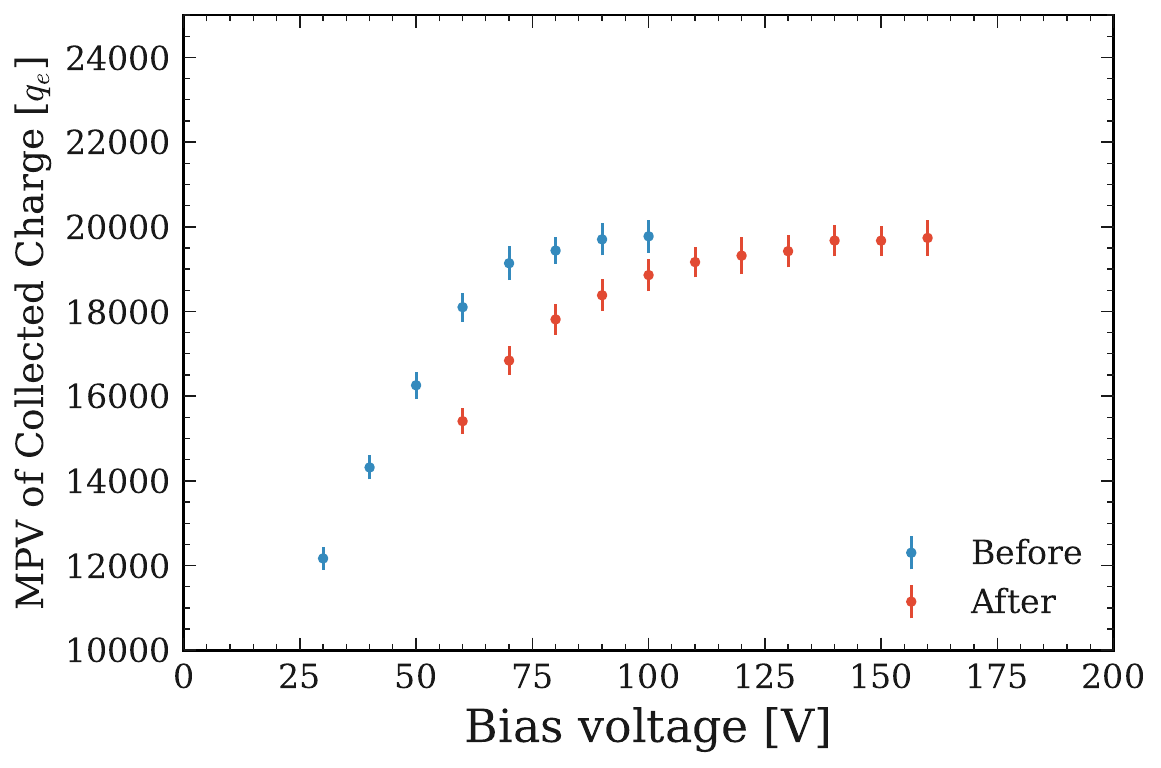}
    \caption{{MPVs} of the signal charge distribution as a function of the bias voltage before (blue) and after (red) irradiation with the {1~MeV neutron equivalent fluence} of ${3.1}\times10^{13}~\si{\cm^{-2}}$.
    The mean and standard deviation of the {50-subset} results are shown by a dot with error bars.}.
    \label{fig:CollectedCharge}
\end{figure}

Figure~\ref{fig:CollectedCharge} displays the MPV variation as the bias voltage increases.
In this figure, the mean and deviation of the 50-subset results are represented by dots with error bars.
Note that we do not show the MPVs for {bias voltages} below \SI{20}{V} before irradiation and below \SI{50}{V} after irradiation because the small signal-to-noise ratios result in too few events for analysis. 
The MPV increases as the bias voltage increases until {it reaches to a plateau at around \SI{70}{V} and \SI{110}{V} before and after irradiation, respectively.}
{The full depletion voltage values were estimated from the $I^2$-$V$ curves to be approximately \SI{70}{V} and \SI{51}{V} before and after irradiation, respectively. A significant deficit in signal charge was observed after irradiation, which is considered to be the result of inefficient charge collection. Although the noise value becomes almost constant above the estimated full depletion voltage, a higher bias voltage is required to recover the MPV after irradiation.}
The plateau MPV is approximately 20000~{$q_e$} both before and after the irradiation. Taking into account the large uncertainty of about 15\% in the APV25 gain calibration, the measured value is compatible with that expected from a \SI{320}{\um} thick silicon sensor ($\sim 24000~{q_e}$)~\cite{Belle-IISVD:2022upf}.
{Average of the MPV ratios between before and after irradiation for all 50 subsets of strips is consistent with unity (0.998 ± 0.002), indicating that there is no significant decrease in the charge collection efficiency due to irradiation.}

\section{Conclusion}
\label{sec:conclusion}
We conducted an irradiation test with \SI{90}{MeV} electron beams on two large Belle~II SVD sensors and ten mini sensors. 
The irradiation was throughout the entire sensor area for the mini sensors, whereas the large sensors were exposed within a \SI{1.6}{\cm} width region parallel to the P-strips near the center of the sensor. 

We estimated the full depletion voltage of the sensors through $I$-$V$ measurement; type inversion of the sensor bulk was observed {at the {1~MeV neutron equivalent fluence} of ${(6\mbox{-}7)}\times 10^{12}$ \si{cm^{-2}}.} {The full depletion voltage increased to about \SI{160}{\volt} at ${3.1}\times 10^{13}$ \si{cm^{-2}}, which approximately corresponds to ten years of the full-luminosity Belle~II operation with a safety factor greater than three. The increase of the depletion voltage is acceptable for the power supply system.}
In addition, we measured an increase in the leakage current proportional to {the 1~MeV neutron equivalent fluence}.
From the linear dependence, we determined the damage constant as {${4.9}\times 10^{-17}$ A/\cm at \SI{20}{\degC}} right after irradiation. 
Furthermore, we tracked the time evolution of the damage constant after irradiation. 
Due to annealing, the damage constant
dropped significantly to approximately 40\% of the initial value in 200 hours and stabilized at approximately 30\% in 1000 hours.
{It is also confirmed that the leakage current of the installed SVD sensors after radiation damage corresponding to a fluence of \SI{3.1e13}{\centi\meter^{-2}}, as estimated from the results of this study while considering possible operating temperature of the SVD sensors, remains acceptable for the power supply system.}

We perform measurements of
noise and signal charge for a large sensor irradiated {up to ${3.1}\times 10^{13}$ \si{cm^{-2}}.}
Sensor noise increased by about 44\% after irradiation, while signal charge was unchanged when measured with a sufficiently high bias voltage {above \SI{140}{\volt}} {and no degradation of the signal charge collection performance was observed.}
{Although the seed charge and noise distributions in the installed SVD sensors depend on factors including the sensor position and strip pitch, an analysis of the measured distributions indicates that the 44\% degradation in the SNR is expected to reduce the cluster detection efficiency by about 3\% only for a subset of the sensors in the innermost layer, while having a negligible impact on the remaining sensors.
It is worth noting that the damage constants independently measured in the installed SVD sensors, reaching up to about $10\times10^{-17}~{\rm A/cm}$, differ from the value obtained in this study. Therefore, interpreting the present result in the context of the actual operating environment involves several sources of uncertainty, including the operating temperature. Nevertheless, these uncertainties are covered by the adopted safety factor.
The results of this study confirm that the SVD sensors will maintain acceptable performance,}
even after irradiation corresponding to approximately ten years of full-luminosity operation.

\section*{Acknowledgement}
The authors thank the ELPH team at Tohoku University, which was reorganized into RARiS in 2024, for delivering high-quality electron beams.
This project has received funding from the European Union's Horizon 2020 research and innovation programme under Grant Agreements No 654168 and 101004761, under Marie Sklodowska-Curie grant agreements No 644294, 822070, 101026516 and 101183137, and ERC grant agreement No 819127. This work is supported by MEXT, WPI, and JSPS (Japan); BMFWF (Austria); CNRS/IN2P3 (France); DAE and DST (India); INFN (Italy); NRF and RSRI (Korea); and MNiSW (Poland).

\newpage

\appendix
\section{Estimation of the full depletion voltage in the low voltage region} \label{sec:logI2}
\setcounter{figure}{0}
When the full depletion voltage {$V_\text{FD}$} is low, the leakage current increases steeply in response to an increase in the bias voltage. 
In this condition, the second-derivative method does not correctly estimate the $V_\text{FD}$ value.

For samples with an expected $V_\text{FD}$ below \SI{20}{V}, we used a different estimation method based on the first derivative of the $I^2$-$V$ curve normalized by the square of the leakage current, $(1/I^2)(\dd I^2/\dd V)$. 

\begin{figure}[htpb]
    \centering
    \includegraphics[width=\linewidth]{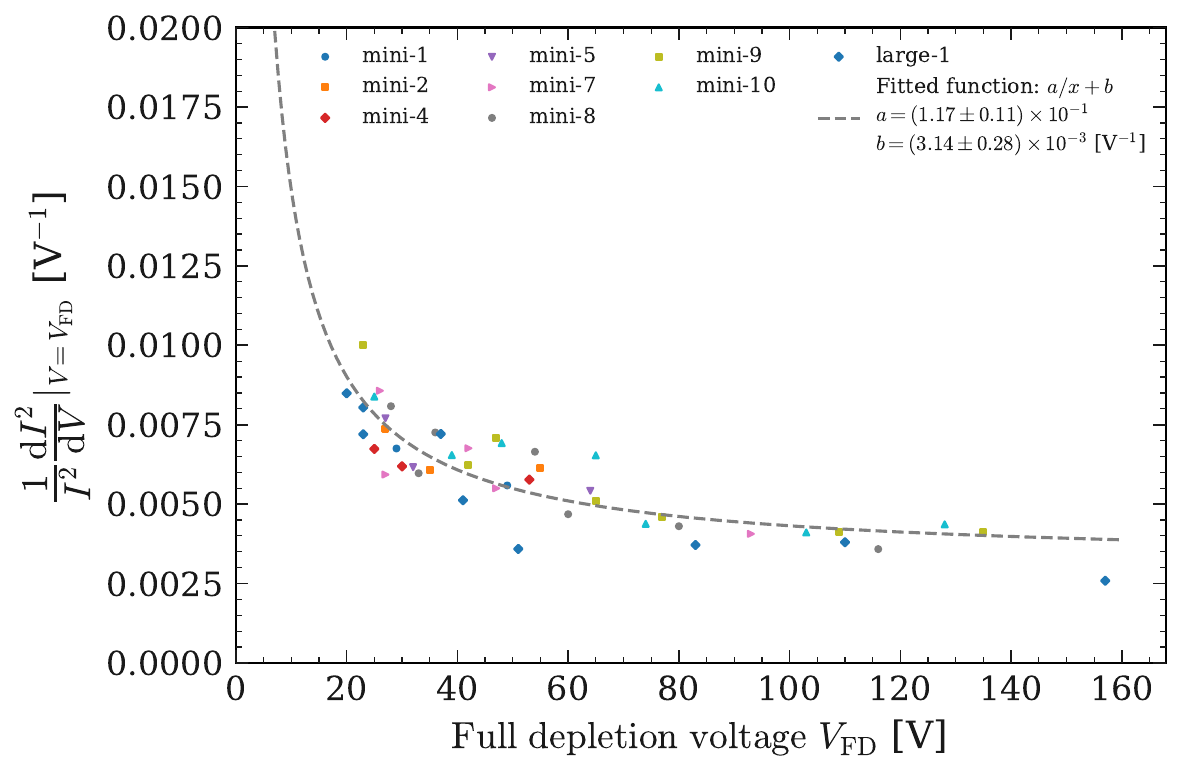}
    \caption{
    {Points indicate the correspondence between the determined $V_\text{FD}$ for the large and mini sensors based on the second-derivative method and the $(1/I^2)({\rm d}I^2/{\rm d}V)$ value at $V\!=\!V_\text{FD}$. The dashed line indicates the fit results of an empirical function $f(V_\text{FD}) = a/V_\text{FD} + b$ to the point distribution, where the parameters assume the values $a = {(1.17\pm0.11)\times10^{-1}}$ and $b = (3.14{\pm0.28})\times10^{-3}~\text{V}^{-1}$.}
    }
    \label{fig:NormalizedSlope}
\end{figure}

Each point in Figure~\ref{fig:NormalizedSlope} shows the correspondence between the $V_\text{FD}$ value determined for the large and mini sensors based on the second-derivative method, i.e., for those sensors and irradiation doses for which this method gives $V_\text{FD}>20$~V, and the value of $(1/I^2)(\dd I^2/\dd V)$ at $V\!=\!V_\text{FD}$.
The point distribution is fitted with an empirical function $f(V_\text{FD})=a/V_\text{FD}+b$, with free parameters $a$ and $b$, where $a$ is dimensionless while $b$ has dimension of 1/voltage; Fig.~\ref{fig:NormalizedSlope} shows the fit result as a dashed line.

We use the $f(V_\text{FD})$ curve extrapolated below 20~V to predict the $(1/I^2)(\dd I^2/\dd V)$ value at $V_\text{FD}$ {when $V_\text{FD}<20~$V}.
In the new method, we calculate the normalized first derivative {of $I^2$}, $(1/I^2)(\dd I^2/\dd V)$, as a function of the applied bias voltage $V$, then compare it with $f(V)$.  We take the bias voltage value that gives $(1/I^2)(\dd I^2/\dd V) = f(V)$ as $V_\text{FD}$.

\begin{figure}[htpb]
    \centering
    \includegraphics[width=\linewidth]{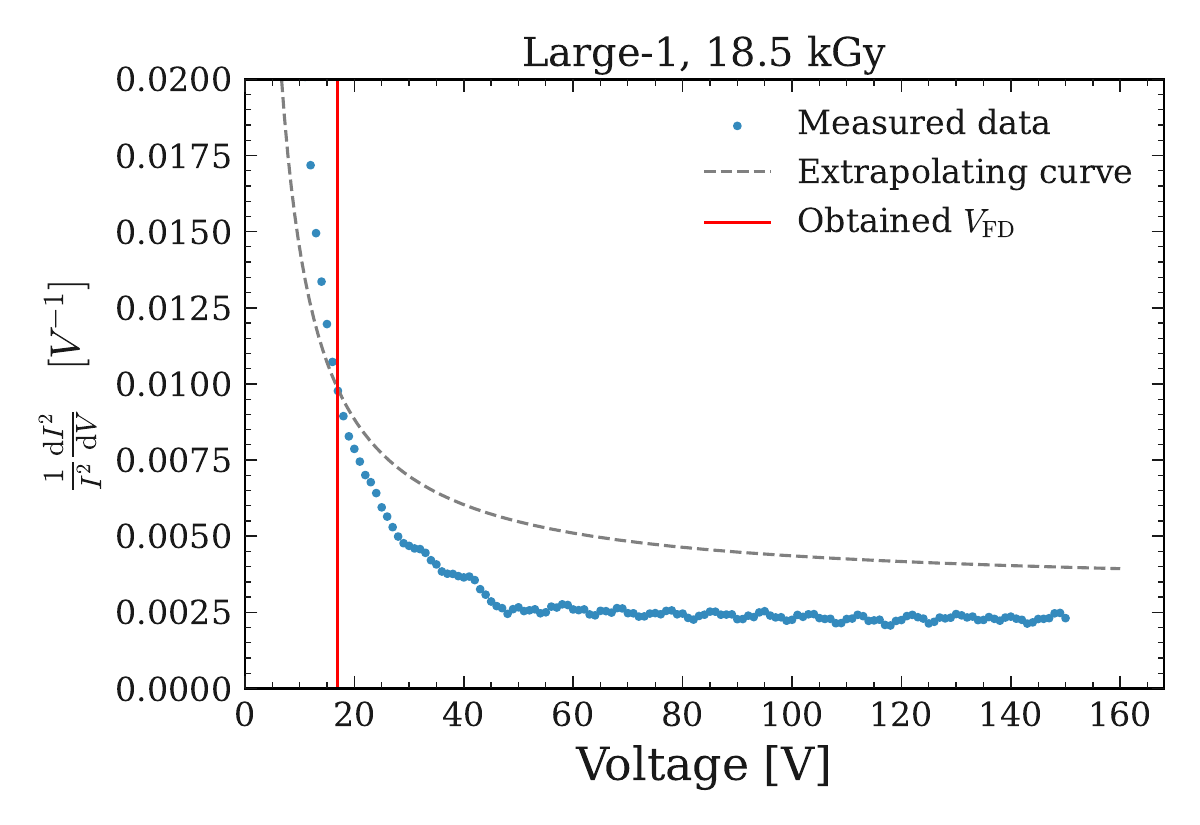}
    \caption{
    {Dots indicate the values of $(1/I^2)(\dd I^2/\dd V)$ calculated at all points of the $I$-$V$ measurement for the large-1 sensor after irradiation at 18.5~kGy. The dashed line is the curve $f(V) = a/V + b$ fitting the points of Figure~\ref{fig:NormalizedSlope}.
    The obtained $V_\text{FD}$ value is {17}~V.}
    }
    \label{fig:Example}
\end{figure}

As an example of the procedure applied to a particular case of low full depletion voltage, Figure~\ref{fig:Example} shows $f(V)$ and the values of $(1/I^2)(\dd I^2/\dd V)$ calculated from the measured $I$-$V$ points: the intersection point between the two curves defines $V_\text{FD}$.

\bibliographystyle{elsarticle-num}
\bibliography{ref}
\end{document}